\documentclass[]{spie} 
 
\usepackage{amsmath,amsfonts,amssymb}
\usepackage{graphicx}
\usepackage[colorlinks=true, allcolors=blue]{hyperref}

\usepackage[]{graphicx}
\usepackage{subcaption}

\title{An adaptive optics upgrade for the Automated Planet Finder Telescope using an adaptive secondary mirror}

\author[a]{Rachel Bowens-Rubin}
\author[b]{Arjo Bos}
\author[a,c]{Philip Hinz}
\author[c]{Bradford Holden}
\author[c]{Matt Radovan}

\affil[a]{University of California Santa Cruz, 1156 High St, Santa Cruz CA, USA}
\affil[b]{TNO Space \& Scientific Instrumentation, Delft, The Netherlands}
\affil[c]{University of California Observatories, 1156 High St, Santa Cruz CA, USA}

\authorinfo{Further author information: (Send correspondence to R. Bowens-Rubin)\\R.B.R.: E-mail: rbowru@ucsc.edu}

\begin{document} 
\maketitle

\begin{abstract}
As we enter the era of TESS and JWST, instrumentation that can carry out radial velocity measurements of exoplanet systems is in high demand.   We will address this demand by upgrading the UC Lick Observatory’s 2.4-meter Automated Planet Finder (APF) telescope with an adaptive optics (AO) system.
The AO upgrade will be directly integrated into the APF telescope by replacing the telescope’s static secondary mirror with a 61-actuator adaptive secondary mirror (ASM) to minimize the disturbance to the spectrograph optics. This upgrade is enabled by the Netherlands Organization for Applied Scientific Research's (TNO) large-format deformable mirror technology, which is constructed using a new style of high-efficiency   hybrid-variable reluctance actuator.
We outline the technical design and manufacturing plan for the proposed APF AO upgrade and simulate the improvement to the  science yield using \textit{HCIpy}.  Our simulations predict the AO upgrade will reduce the PSF instabilities due to atmospheric turbulence, concentrating the light on the spectrograph slit by a multiplicative factor of more than two (doubling the telescope’s observing efficiency) for targets as dim as I = 14. 
When completed, the APF adaptive secondary mirror will be among the first pairings of an ASM with a radial velocity spectrograph and become a pathfinder for similar AO systems in telescopes of all sizes. 
\end{abstract}


\section{INTRODUCTION}
\label{sec:intro}  

Measuring radial velocity (RV) is one of the most successful methods for detecting and characterizing new exoplanets.  Not only have RV surveys revealed hundreds of new planets, RV measurements can be used to constrain an exoplanet's mass, bulk density, and orbital parameters.  
The need for RV instruments is expected to grow over the next decade as exoplanets identified by the Transiting Exoplanet Satellite Survey (TESS) become followup targets for transit spectroscopy using the James Webb Space Telescope (JWST) and require precision orbital ephemeris measurements.~\cite{Plavchan2015}
NASA identifies the need for increased RV capabilities in low and high precision in the Exoplanet Exploration Program Gap List (Section SCI-09)~\cite{Stapelfeldt01} stating, ``There are insufficient precision RV resources available to the community to follow up all K2 and TESS candidates that may be relevant to spectroscopic studies with JWST and ARIEL.''  

A cost-effective option for addressing the RV-instrument shortage is to improve the performance of our current on-sky instrumentation. 
Adaptive optics has the potential to improve (1) the observational efficiency by concentrating the point spread function of the telescope and reducing the centroid variation of the PSF, and (2) increase the instrumental stability by
creating a PSF that varies less than is possible with a seeing-limited telescope.
The biggest source of noise in RV measurements is point spread function (PSF) instability.~\cite{Spronck2015}  While we have approaches to fit and remove the effects from PSF variations, it requires high signal-to-noise in the spectra which adds to the integration time for each observation.  
Instruments such as iLocater~\cite{Crepp2016}, PARVI~\cite{Gibson2020} and HISPEC~\cite{Mawet2019} are exploring the use of diffraction-limited fiber-injection, however these approaches require high performance AO to preserve reasonable efficiency through the single mode fiber.  When an adaptive optics system is combined with a slit spectrograph in contrast, minor corrections to the image stability and PSF size can lead to dramatic throughput improvements if the corrections are on the order of the width of the slit. 
Any throughput improvements will lead to additional photon collecting power which can be used to (1) reduce the necessary  integration time to reach the same precision allowing for more targets to be observed per night, (2) access dimmer targets, reducing the demands on RV instruments at larger telescopes, or (3) achieve a higher SNR spectra using the same exposure to better remove the effects of the PSF variations and improve the precision of the RV measurement.

While adaptive optics systems have the potential to provide large gains for radial velocity spectrographs, AO has not traditionally been integrated into spectroscopic instruments because of the additional expense and complexity. Over the past five years, gains in large-format deformable mirror technology have made it significantly less costly to replace the secondary mirror of the telescope with an adaptive secondary mirror (ASM). Using an ASM eliminates the need to integrate the AO system near each science instrument and instead allows the deformable mirror and wavefront sensing path to exist as part of the telescope optics. This allows for minimal design modifications to science instruments and makes the AO system accessible to any science instrument on the telescope.

In this paper, we overview the initial designs for the adaptive optics upgrade of the Automated Planet Finder Telescope (APF) located at Lick Observatory in Mount Hamilton, California that includes an adaptive secondary mirror. 
The APF ~\cite{Vogt2014} is a 2.4m robotic telescope with an integrated visible light spectrometer that has been operational since 2013.  It is a dedicated exoplanet radial velocity facility that provides an ideal testbed for demonstrating AO-stabilized radial velocity detection.  
This system will be among the first AO-stabilized RV spectrometers and will become the smallest ASM ever tested on-sky.   If proven successful, this project could be a pathfinder for improving both AO and RV instrumentation for telescopes of all sizes.

\section{ONGOING AND FUTURE SCIENCE AT THE APF}

The APF science program has two core goals: 
(1) follow-up known short period exoplanet discoveries with intense, high cadence observation and (2) discover new exoplanets through multi-year monitoring.
Both of these programs utilize the key advantage of the APF, it is a dedicated facility with a single instrument. 

The combination of high cadence and multiple years of data allows complicated systems to be disentangled and on-going observations will continue to yield further dynamically interesting systems, 
including, Vogt et al 2015~\cite{Vogt2015}, Vogt et al 2017~\cite{Vogt2017}, Rosenthal et al 2019AJ~\cite{Rosenthal2019AJ}, and Lubin et al 2021~\cite{Lubin2021}. 
To date, the APF-50 survey~\cite{Fulton2015} has been the most successful, using the ability to repeatedly observe a system to beat down the noise and find multiple Neptune sized systems in volume-limited surveys~\cite{Fulton2017} resulting a large public data releases~\cite{Rosenthal2021}. 
For all these systems, 
it is possible to obtain a better measurement with hundreds of multiple-epochs, low-SNR observations  than fewer worse-cadence observations on a facility with better individual measurements~\cite{Fulton2015}.

The value of the APF was demonstrated during the Kepler era, particularly with the K2 mission.  With K2, a larger number of transiting systems were found around nearby, bright stars that are accessible to the APF. The APF’s ability to rapidly change targets (and target within a day of the announcement) means that the APF has been important in following up K2 systems. The high-cadence, no-interruption observations meant that the APF played a critical role in understanding HD3167~\cite{Christiansen2017}, a triple planetary system with two transiting planets and one non-transiting planet. Because other facilities are often scheduled around the moon or cannot achieve daily cadence, the APF velocities were critical in disentangling the various signals from sample aliasing, allowing the masses of the transiting planets to be measured. The TESS era has further borne this out, with the discovery of an exoplanet through radial velocities that was only later found to be transiting~\cite{Scarsdale2021}. 

The daily cadence and long-term monitoring are critical for the current science programs. By leveraging APF’s excellent cadence concurrently with TESS observations, astrophysical RV noise from stellar activity may be optimally modeled and subtracted~\cite{Aigrain2012}, revealing low-mass exoplanets that APF would not normally be sensitive to. The automated queue-scheduling nature of APF makes it an ideal facility for long-term monitoring of bright host stars with known planetary systems. One such project is monitoring a collection of such systems with known planets  that also show evidence for possible long-period companions~\cite{Hurt2021,Lubin2021}. The APF observations will track the linear trends present in the RVs and help to understand the architectures and origins of systems with planets as the time baseline is increased.

Recent studies have uncovered evidence of a radius dichotomy for planets with $R < 4  R_{Earth}$ among the Kepler planet population~\cite{Fulton2017AJ}. To understand the cause of this radius dichotomy, it is crucial to obtain both radii and masses for a population of planets spanning the 1.5-2  $R_{Earth}$ radius range. If a corresponding mass dichotomy exists among these planets, then the observed radius gap is likely acting as a tracer of planet formation and inherent composition.  If we see no such mass split, then the observed radius gap is likely tracing photo-evaporation as some subset of these planets lose their atmospheres later in life. 

The best hope for further populating the small planet mass-radius diagram in the near term lies with measuring the mass of exoplanets discovered by TESS. TESS began science operations in July 2018, and is searching for small, transiting planets around the closest, brightest stars - exactly the population required to answer this key question in planet evolution~\cite{Ricker2015}.  The APF's ability to measure nightly velocities has already contributed to measuring exoplanet masses for planets with $R < 4 R_{Earth}$ that were detected by TESS (for example; Scarsdale et al 2021 ~\cite{Scarsdale2021}, Hurt et al 2021 \cite{Hurt2021}, Heidari et al 2021 \cite{Heidari2021}). The exoplanet discovered in Heidari et al 2021~\cite{Heidari2021} is uniquely interesting, as it has an Earth-like density ($\sim5-5.5$ gm/cm$^3$) despite having a Neptune-like mass ($\sim$14-16 M$_{Earth}$).  This exo-Neptune imparts a velocity on the host star HD~207897 of $\sim4.2-4.6$ m/s, only slightly larger than the 3 m/s errors on the velocity measurements. Simply improving the effective seeing of the APF will increase the number of stars with TESS detections which can be surveyed on a nightly basis by increasing the signal to noise in a fixed exposure time, halving the 1600 seconds exposures required for HD207897. The real advancement would be from improving the stability of the input beam into the spectrometer, which will  increase the precision at a fixed signal noise. That could lower the 3 m/s errors to $1-2$ m/s as seen in Spronck et al 2015~\cite{Spronck2015}.

The APF provides a unique long-baseline data set that can be revisited to search for long period companions. Some of the best studies of the occurrence rates of long period Jupiter-massed exoplanets from radial velocities have utilized the long-term programs from both the Lick and Keck Observatory~\cite{Cumming2008,Bryan2016,Rosenthal2021,Fulton2021}. The ability to perform these long-baseline surveys on host stars that already have TESS detected exoplanets provides key indicators for the formation and evolution of exoplanets. Bryan et al 2018 ~\cite{Bryan2018} found that super-Earths are significantly more likely to be in systems with Jupiter-mass, long-period planets. Lubin et al 2021~\cite{Lubin2021} revealed a Saturn-massed long period companion in a system with three TESS-detected super-Earths. 
By constructing large samples of TESS detections with long period follow-up, the data from the APF can be used to understand the interplay between the typical short period super-Earth so commonly found with Kepler and TESS, and larger mass planets on longer period orbits. 

Finally, the APF can be used to determine the masses and periods of exoplanets discovered with a single transit. In Dalba et al 2020~\cite{Dalba2020}, the authors measure HD~332231 with the APF, HIRES, and the SONG telescopes. The RV data allowed the prediction of the time of the next transit. Many single transit detections from TESS will have no subsequent data within the TESS lifetime. Radial velocity data are the only way to measure these exoplanet's orbital properties, as well as their masses. By measuring the RV of single-transit TESS candidates like this, we provide a sample of exoplanets that have radii measurements (from the single TESS transit) but have a longer period than the exoplanets detected by TESS data alone.
This is essential for providing a second key set of data for understanding the formation of exoplanet system and their distribution of mass, radii, and compositions as a function of orbital parameters.

\section{CURRENT TECHNOLOGY}
\subsection{The Automated Planet Finder Telescope}

\begin{figure}[h!]
\centering
\begin{subfigure}{0.49\textwidth}
  \centering
  \includegraphics[width=1.0\linewidth]{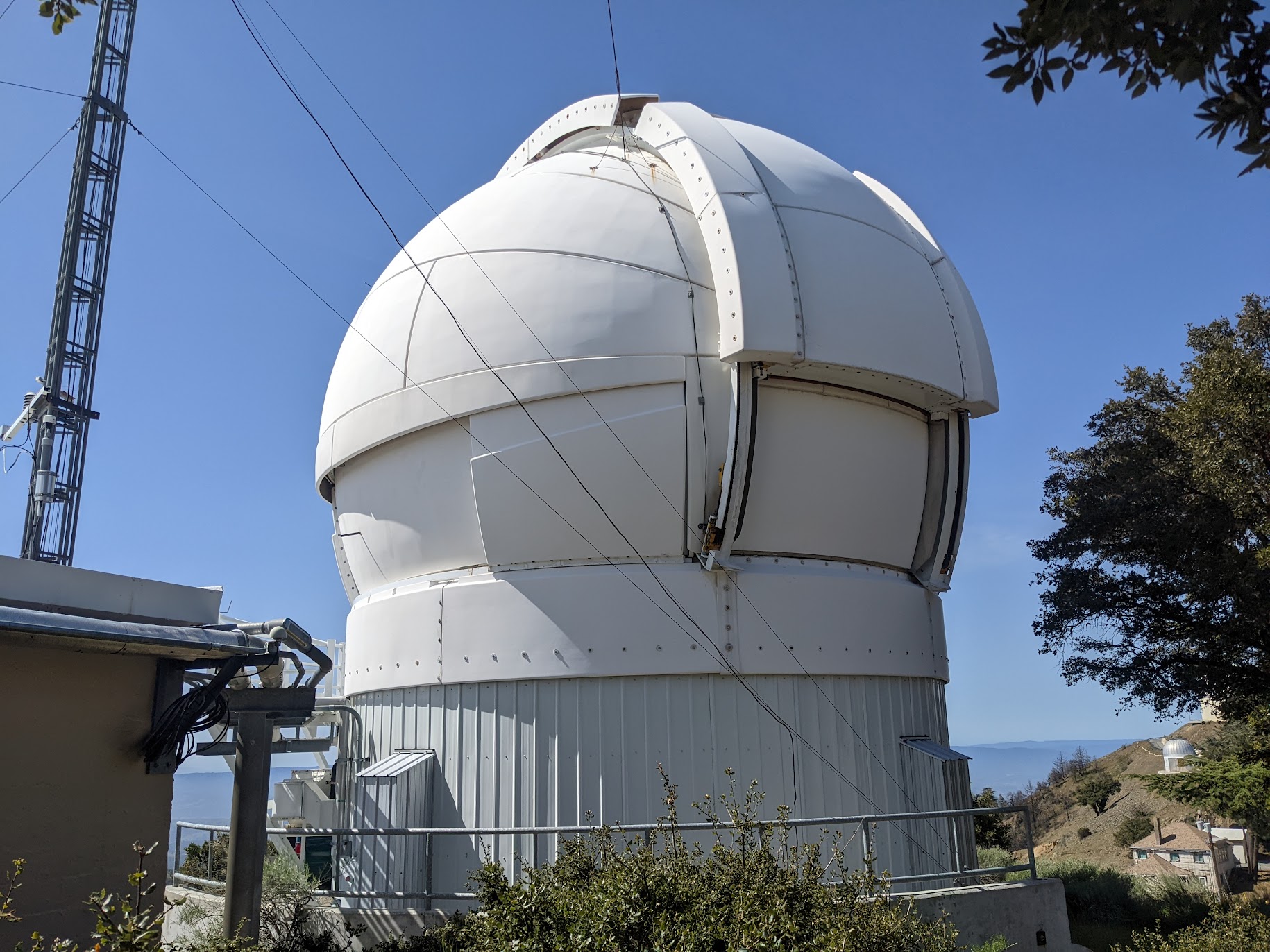}
\end{subfigure}
\begin{subfigure}{0.49\textwidth}
  \centering
  \includegraphics[width=1.0\linewidth]{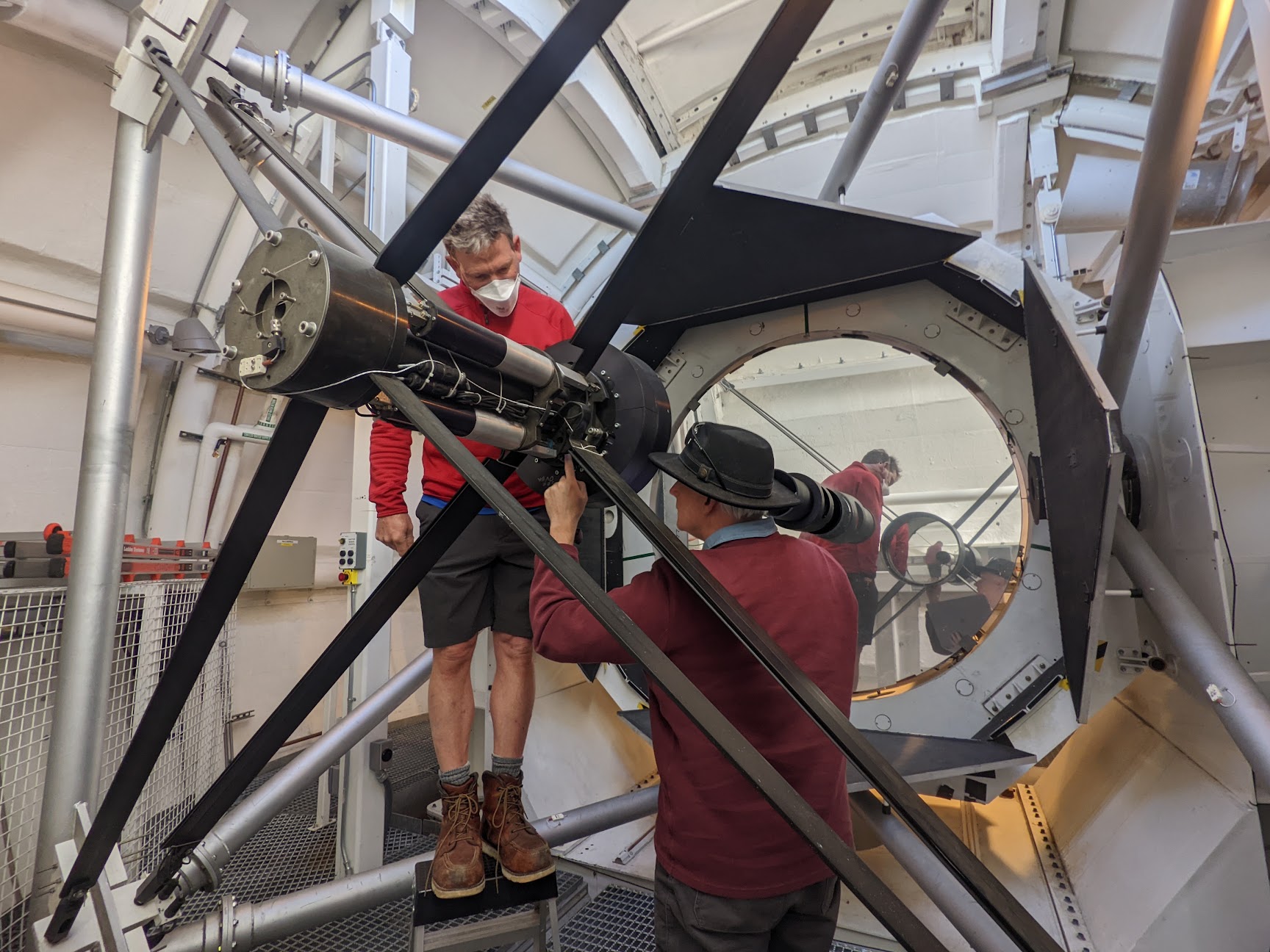}
 \end{subfigure}
 \caption[example]{\textbf{The 2.4m Automated Planet Finder Telescope} \label{fig:APFpic} 
 }
   \end{figure}

The Automated Planet Finder (APF) is a 2.4-meter telescope located at the University of California Lick Observatory near San Jose, California that operates fully robotically (Figure \ref{fig:APFpic}). 
An overview of the basic properties of the APF and observing site are listed in Table \ref{tab:APFoverview}.  A typical value for the seeing at Lick Observatory is 1.5 arcsec with tip/tilt motions of 0.2 arcsec. 

The APF was constructed as a dedicated facility to search for extrasolar planets using an optical echelle spectrometer and an iodine gas absorption cell. The spectrometer is optimized to take precision radial velocity measurements
around nearby low-mass dwarf stars.
~\cite{Burt2014}
Such cooler stars are rich in stellar lines in the iodine region. That high information density combined with the high resolution of the Levy spectrometer yields up to a 1 m/s precision for radial velocities.~\cite{Burt2016}
 The APF is optimized for exoplanet science using a spectrometer with spectral resolving power of 
R$\approx$100,000. For typical observations, the APF uses an image plane slit width of 0.5 arcsec to optimize the achievable spectral resolution.

The APF optical system and the measured on-sky performance is documented on the \href{https://apf.ucolick.org/intro.html}{APF UCO Lick website}, in the  
\href{https://apf.ucolick.org/docs/APF_prospectus_02july2013.pdf}{APF prospectus from July 2013}, and in Jennifer Burt's PhD dissertation:  \href{https://escholarship.org/uc/item/8c32w845}{The Automated Planet Finder telescope's automation and first three years of planet detections}\cite{Burt2016}.


\begin{table}[]
\begin{center}
\caption{\textbf{Overview of the Current APF Telescope and Performance} \label{tab:APFoverview}}    
\begin{tabular}{|l|l|}
\hline
Telescope size                                                                              & 2.4 m                                                              \\ \hline
Optical prescription                                                                        & f/15                                                               \\ \hline
Location                                                                                    & \begin{tabular}[c]{@{}l@{}}Mt Hamilton, CA\end{tabular} \\ \hline
Typical seeing at site                                                                      & 1.5 arcsec$^{[*]}$                                                         \\ \hline
Type of Spectrograph                                                                        & Slit Spectrometer$^{[*]}$                                                  \\ \hline
Wavelength Reference                                                                        & Iodine gas cell in beam$^{[*]}$                                                  \\ \hline

Slit Width Options                                                                          & 0.5-2"$^{[*]}$                                                           \\ \hline
Spectrograph Resolution                                                                     & R=110,000 (0.5"), R=80,000 (1")$^{[*]}$                                    \\ \hline
Science spectral range                                                                      & 374 - 680 nm$^{[*]}$                                                       \\ \hline
\begin{tabular}[c]{@{}l@{}}Magnitude Limit of \\ the Guide Camera\end{tabular}              & V = 4 - 15$^{[\dagger]}$                                                         \\ \hline
System throughput                                                                   & $21 \pm 4\%^{[\dagger]}$                                                         \\ \hline
\begin{tabular}[c]{@{}l@{}}Median RV Precision \\ (meas. July 2013-March 2016, V=4.68))\end{tabular} &  1 m/s$^{[\dagger]}$                                                       \\ \hline
\end{tabular}
\newline
$[*]$ Vogt et al 2014\cite{Vogt2014}, $[\dagger]$ Burt et al 2016\cite{Burt2016}
\end{center}
\end{table}

\subsection{TNO Large-Format Deformable Development}
The proposed APF adaptive secondary mirror relies on a large format deformable mirror technology in development by the Netherlands Organization for Applied Scientific Research (TNO). TNO has developed a new style of hybrid-variable reluctance actuator to be used in  large-format deformable mirrors.  The current implementation of this actuator design can fit into a smaller volume and is $\sim$75 times more efficient than the Microgate generation two voice-coil actuators (HVR actuator efficiency~\cite{Kuiper2018} $= 38 N/\sqrt{W}$; MMT and LBT actuator efficiency~\cite{Riccardi2003} $= 0.5 N/\sqrt{W}$).
This technological breakthrough is in the early stages of providing an opportunity for the astronomy community to integrate adaptive secondary mirrors into telescopes of all sizes, without needing complex and costly cooling systems. 

A 57-actuator large-format deformable mirror made using TNO’s 2016 generation of actuators was tested by the UC Santa Cruz Lab for Adaptive Optics in the first half of 2020.  While this mirror was designed for demonstration in a lab setting, the testing verifies that the TNO actuators can have the stroke, linearity, and lack of hysteresis to meet or exceed on-sky needs~\cite{Bowens-Rubin2020}.  

TNO began constructing the first large-format deformable mirror for use on-sky as an adaptive secondary mirror in 2019-20 for the UH-88 telescope located on Maunakea as part of a separately funded NSF ATI program led by Mark Chun.  This adaptive secondary is 63 cm in diameter and is made using 211 actuators.  The actuators are from TNO’s 2020 generation which can accommodate larger actuator spacing and be arranged in a circular grid.  This system is expected to be ready for deployment in the telescope in 2023\cite{Kuiper2020}.

As a companion to the UH-88 adaptive secondary, a 19-actuator lab demonstrator was constructed for the Lab for Adaptive Optics using the same actuator style as the UH-88 ASM.  The 19-actuator mirror is being used to validate the UH-88 ASM technology and guide the engineering development at TNO~\cite{Bowens-Rubin2021}. 
This mirror was also the first to be constructed using a thin, low-cost facesheet that was flattened using the glass slumping method
proposed for the APF ASM.

The APF ASM facility would provide an additional system to retire risks for the TNO technology. This helps with the maturation of the technology for large aperture facilities.  For example, an ASM based on TNO actuators is currently being studied via the Keck community internal white paper process.  A demonstration that these devices are routinely used for science on medium-aperture telescopes such as the APF can provide information about its robustness, performance, and operational parameters. 

The APF ASM will also develop the TNO technology beyond the UH-88 ASM and accompanying lab prototype by exploring the following:

\begin{itemize}
\item Alternative backing structure material (silicon carbide or carbon fiber, the UH-88 structure is aluminum). Silicon carbide and carbon fiber are a closer match to the facesheet coefficient of thermal expansion resulting in less thermal deformation.  They also provide testing of higher specific-stiffness materials as a backing structure for larger ASM’s.  

\item A larger secondary curvature (f/1.6 for APF, vs. f/3.4 for the UH-88).  The APF ASM is similar to optical prescriptions needed for large aperture telescopes and more modern medium aperture telescopes.

\item A more efficient actuator design. TNO is developing an actuator that is smaller, simpler to fabricate, and with a faster response time compared to the UH-88 actuator.
\end{itemize}

\section{DESIGN OF THE APF ADAPTIVE OPTICS UPGRADE}

\subsection{Overview}

\begin{figure}[h!]
\centering
\begin{subfigure}{0.35\textwidth}
  \centering
  \includegraphics[width=1.0\linewidth]{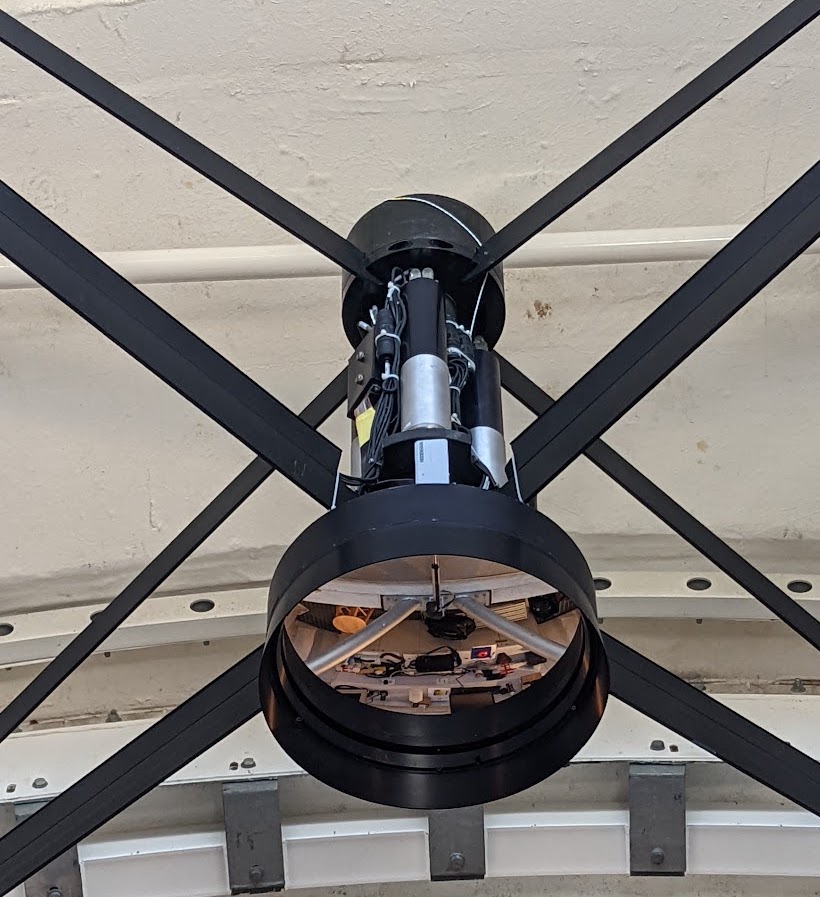}
\end{subfigure}
\begin{subfigure}{0.56\textwidth}
  \centering
  \includegraphics[width=1.0\linewidth]{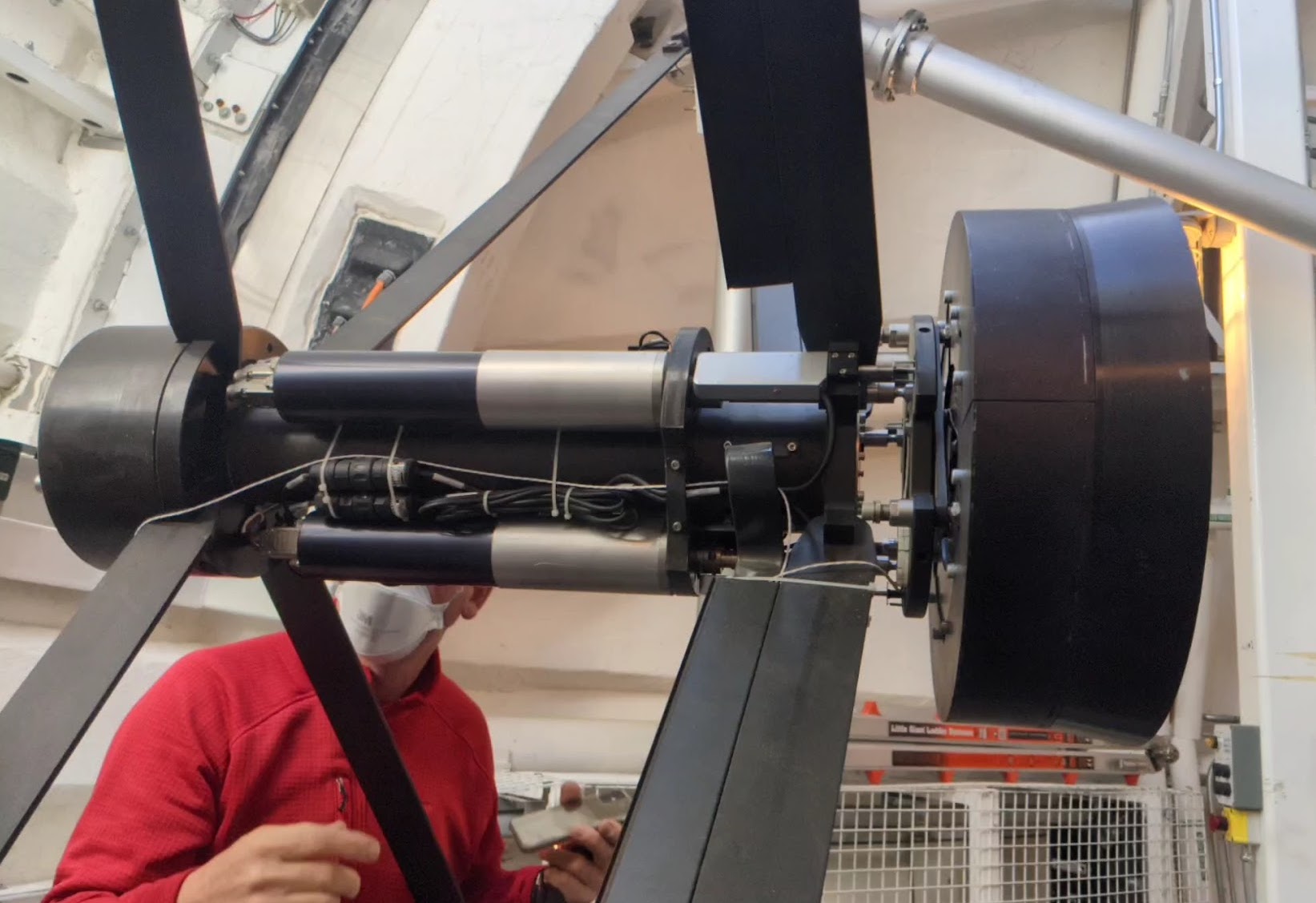}
 \end{subfigure}
 \caption[example]{\textbf{The APF static secondary mirror} mounted on the hexapod that is used to set the focus. \label{fig:apfsecondary} 
 }
   \end{figure}

The design of an AO system to stabilize the APF PSF can be made with two main additions requiring minimal modifications to the telescope. 
The first addition is a replacement of the current static secondary mirror (pictured in Figure \ref{fig:apfsecondary}) with an adaptive secondary mirror (Figure \ref{fig:APFCAD}). The second is an off-the-shelf wavefront sensor and associated optics for the WFS light path (shown in Figure \ref{fig:layout}).

An overview of the proposed AO system is listed in Table 2. The adaptive optics upgrade requires a wavefront sensor that uses light not employed for science observations. A suitable passband is 700-1000 nm, since it is unused in the Levy spectrometer.  Since the spectrometer is observing a bright star, the WFS will be fixed on-axis.  We set the spatial sampling at 8 subapertures across the telescope diameter. This will set both the deformable mirror actuator geometry and Shack-Hartmann lenslet size.  We arrived at this value after exploring different spatial samplings in the AO simulation.  While better samplings improved the Strehl, the coarse slit size resulted in little improvement in throughput above this value. 

RV instruments such as the APF/Levy favor observing bright stars to make these precise measurements.  As such, there are plenty of photons available to measure and correct the wavefront of the light going into the spectrometer.  Further, there is no anisoplanatism effect since the “guide star” used for wavefront control is the science object itself. This allows for a simple implementation of the AO system requiring a natural guide star wavefront sensor on axis. Because the deformable mirror can be integrated into the telescope optics, no new optics are needed beyond a dichroic to extract the light used for the wavefront sensor.

\begin{table}
\begin{center}
\caption{\textbf{Overview of the proposed AO-assisted APF}}
\begin{tabular}{|l|l| l|}
\hline
\bf{Specification}       & \bf{Value}   & \bf{Notes}       \\ \hline
Deformable Mirror Location       & Secondary mirror   &        \\ \hline
Number of Actuators              &   61           & \\ \hline
Wavefront Sensor (WFS)           & Alpao Shack-Hartmann EMCCD& \\ \hline
WFS Bandpass                    & 700-1000nm                 &  \\ \hline
WFS Sampling                 & 8 subapertures across diameter &   \\ \hline
Operating limit with AO &      15 in I band & same as current guider            \\ \hline
Corrected Image Width           &             0.9" & for 1.5" seeing                \\ \hline
Corrected Image Jitter         &                  0.05" & improvement of 4X         \\ \hline
Expected Throughput              &         61\% & improvement of 2.9X                  \\ \hline
Expected Median RV Precision            &           0.5 m/s     &            \\ \hline
\end{tabular}
\end{center}
\end{table}

\subsection{The Adaptive Secondary Mirror}

\begin{figure}[h!]
\centering
\begin{subfigure}{0.65\textwidth}
  \centering
  \includegraphics[width=1.0\linewidth]{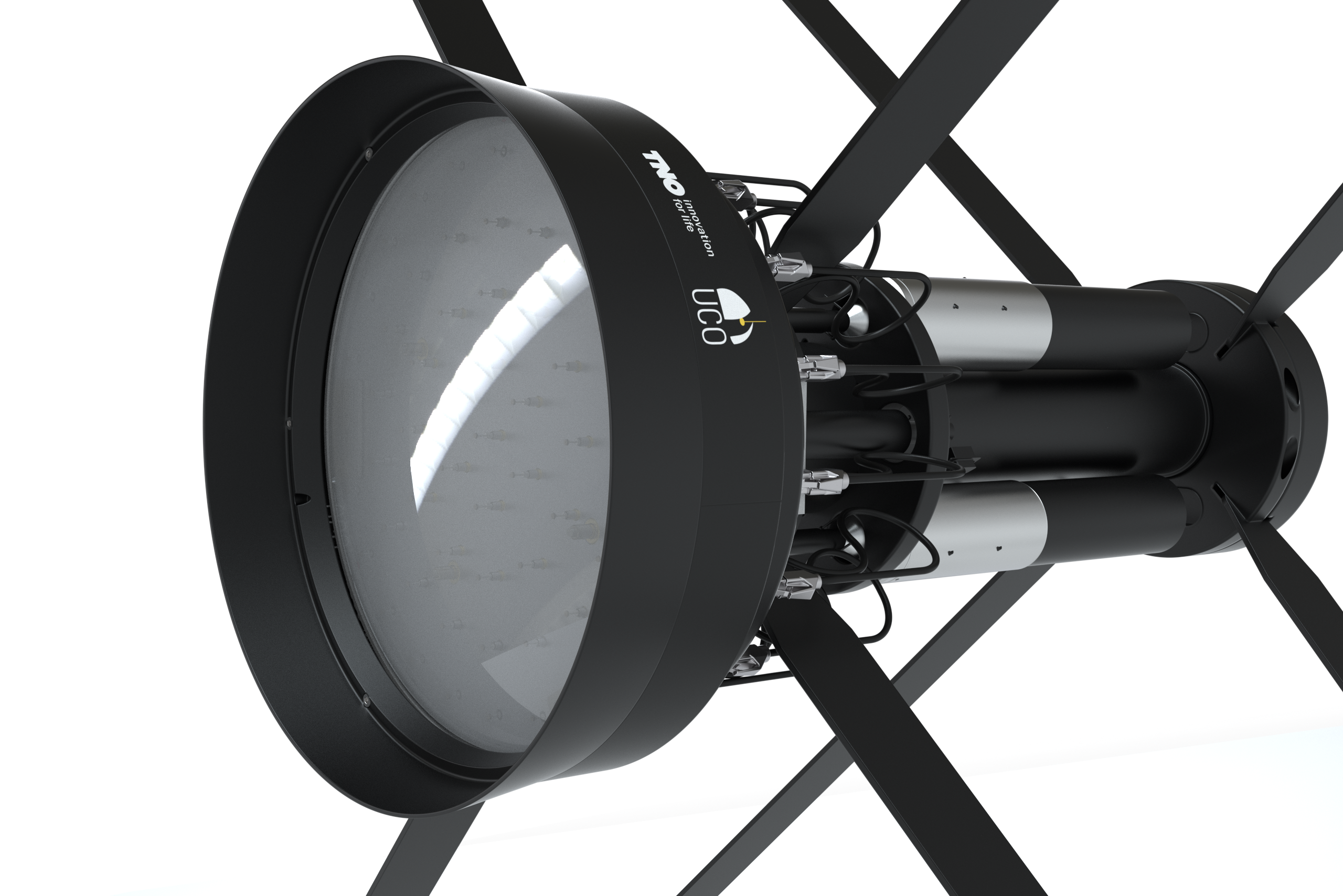}
\end{subfigure}
\begin{subfigure}{0.30\textwidth}
  \centering
  \includegraphics[width=1.0\linewidth]{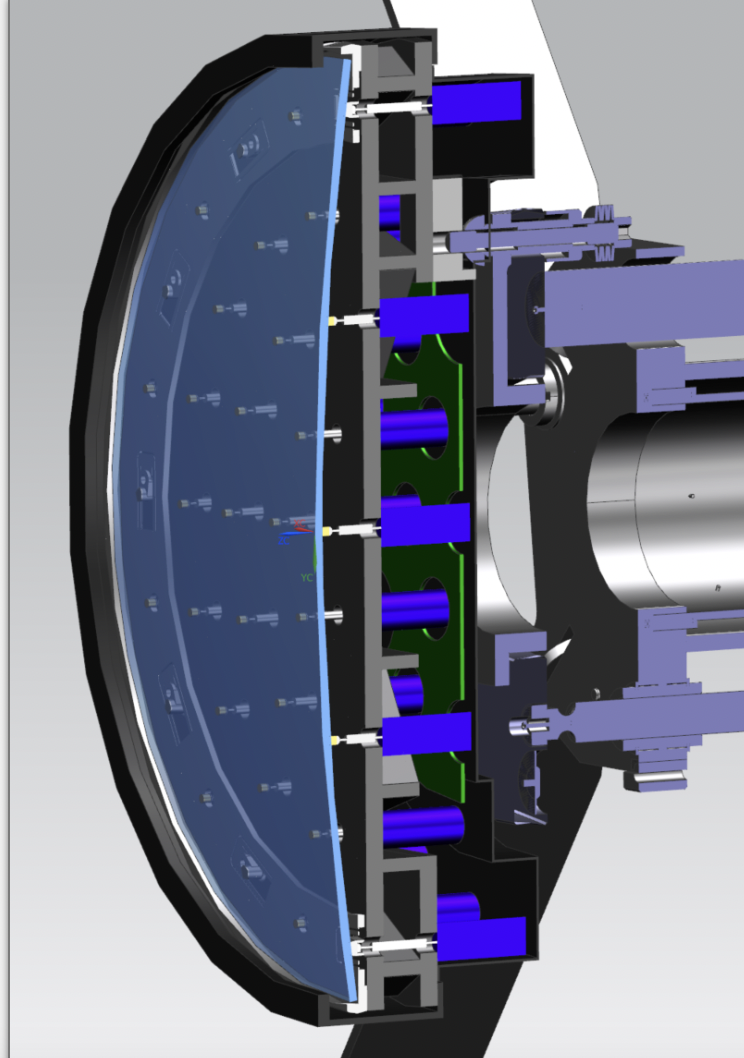}
 \end{subfigure}
 \caption[example]{\textbf{The APF ASM secondary assembly model as designed by TNO. } The ASM has 61 actuators laid out in a hexapolar arrangement.  The ASM will make use of the existing focus and tip-tilt assembly from the current static secondary mirror assembly. \label{fig:APFCAD} 
 }
   \end{figure}

The adaptive secondary mirror for the APF will be manufactured with 61 of TNO's third-generation hybrid variable reluctance actuators, applying a hexapolar actuator pattern with one center actuator. The third-generation actuators are based on the UH-88 ASM actuator design and the prototype deformable mirror being tested in the Laboratory for Adaptive Optics, but optimized for manufacturability and suited for smaller pitch sizes - down to 16 mm. The mirror facesheet will be a 37 cm diameter, convex hyperbola with a radius of curvature of R=1198 mm and a conic constant of -1.49, matching the existing APF optical prescription. A rendered image of the TNO ASM mounted to the top-end is shown in Figure \ref{fig:APFCAD}.  

UC Observatories will provide the curved mirror facesheet as part of its research and development in low-cost, hot forming techniques to shape precision optics. To construct the facesheet, we will use a 65 cm diameter, 3.3 mm thick polished borofloat flat blank placed on top of a ring support. We will heat the blank to glass slumping temperatures under the weight of a top ring sized to be just outside the 37 cm optical diameter. This top weight and the right peak temperature causes the glass to slump into the desired curvature. We have proven this process to create curved shells with the required optical quality form shells with R=1200 mm on 150 mm diameter sheets.  We are currently scaling up this process to make larger diameter facesheets. This hot forming process is more economical than current approaches that polish and then thin the shell to the required thickness.  For a UH-88 size shell, this process has the potential to be 10-30\% the cost of the current state of the art, essentially removing the facesheet as a dominant cost component of an ASM.

For the UH-88, TNO has contracted with the Fraunhofer Institute to make a curved facesheet using more standard hot forming techniques and a solid mandrel.  Should we run into issues scaling up the process, a backup for the facesheet fabrication would be to work with the Fraunhofer group to fabricate a shell for the APF adaptive secondary mirror.

TNO will be responsible for the fabrication of the actuators and backing structure as well as integrating the facesheet with the ASM assembly.  They successfully carried out a  similar integration process for a lab prototype of the UH-88 ASM.

\begin{figure}[h!]
   \begin{center}
   \begin{tabular}{c}
   \includegraphics[height=5.9cm]{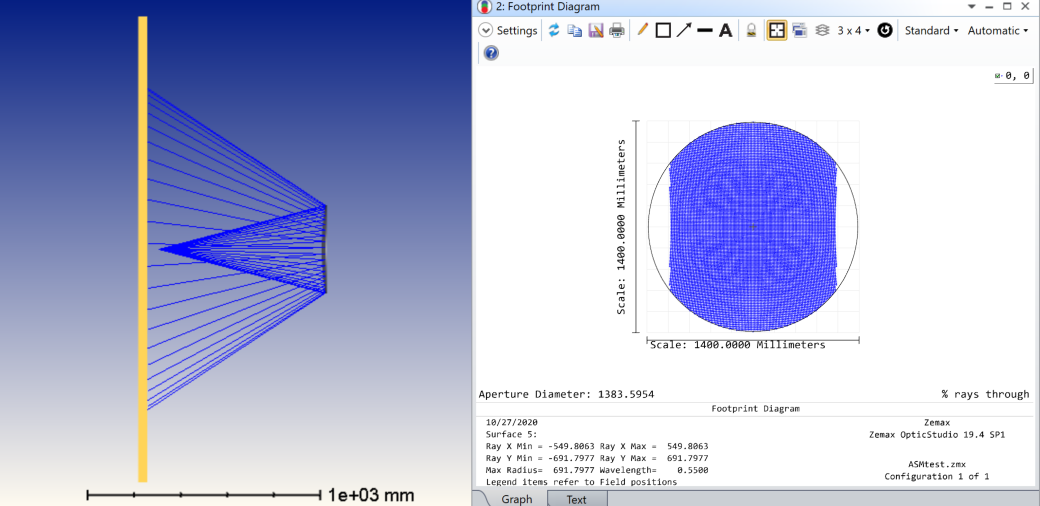}
   \end{tabular}
   \end{center}
   \caption[example] 
   { \label{fig:deflect} 
 \textbf{Layout for a deflectometry test for the APF ASM.}  A large (1.1 m x 2 m) LCD screen will be used to illuminate the ASM with a pattern that will be recorded by a camera.  The screen does not completely illuminate the optic as seen by the camera (right), but can be rotated to measure the complete surface. 
 }
   \end{figure}

Once delivered, the ASM will be tested via a phase-measuring deflectometry test (Figure \ref{fig:deflect}). 
The system will be set up to be illuminated by a large LCD screen.  Horizontal and vertical sinusoidal patterns are displayed on the screen and captured by a camera that views the patterns reflected off the test optic. The resulting measured intensities at each point on the mirror can be used to reconstruct the slopes or deflection imparted by each section of the mirror and integrated to calculate the surface shape.  This test is simple to set up, provides high dynamic range, and can operate without the need for additional large optics.

The ASM will undergo a poke test to measure the influence functions of all the actuators.  Once these are measured, we will have an initial interaction matrix to deform the facesheet.  Further refinements of the interaction matrix will be achieved by applying modal shapes to the mirror to map out achieved versus commanded surface deformations.
Once the deflectometry tests have been completed under standard laboratory conditions, we will carry out the same measurements over a range of elevation angles from zenith to horizon.  UCO has a cold chamber that can carry out the same tests at temperatures down to 0$^{\circ}$C.

\subsection{The Wavefront Sensor}

The wavefront sensor (WFS) for the system will be built around an Alpao standard Shack-Hartmann WFS camera placed on-axis. Because the wavefront sensor requires good sensitivity in the red and good noise performance to be able to operate on the faintest stars for an APF observation, we will use an electron multiplied CCD (EMCCD).  Alpao has developed a Shack-Hartmann EMCCD device with low latency and high framerates suitable for our needs.  The device has 16x16 sub-apertures and can operate at up to 1004 Hz.  Our optical design calls for 8x8 sub-apertures, so we will be able to operate the device at up to 1838 Hz with a frame latency of 69$\mu$s.  

\begin{figure}
   \begin{center}
   \begin{tabular}{c}
   \includegraphics[height=15cm]{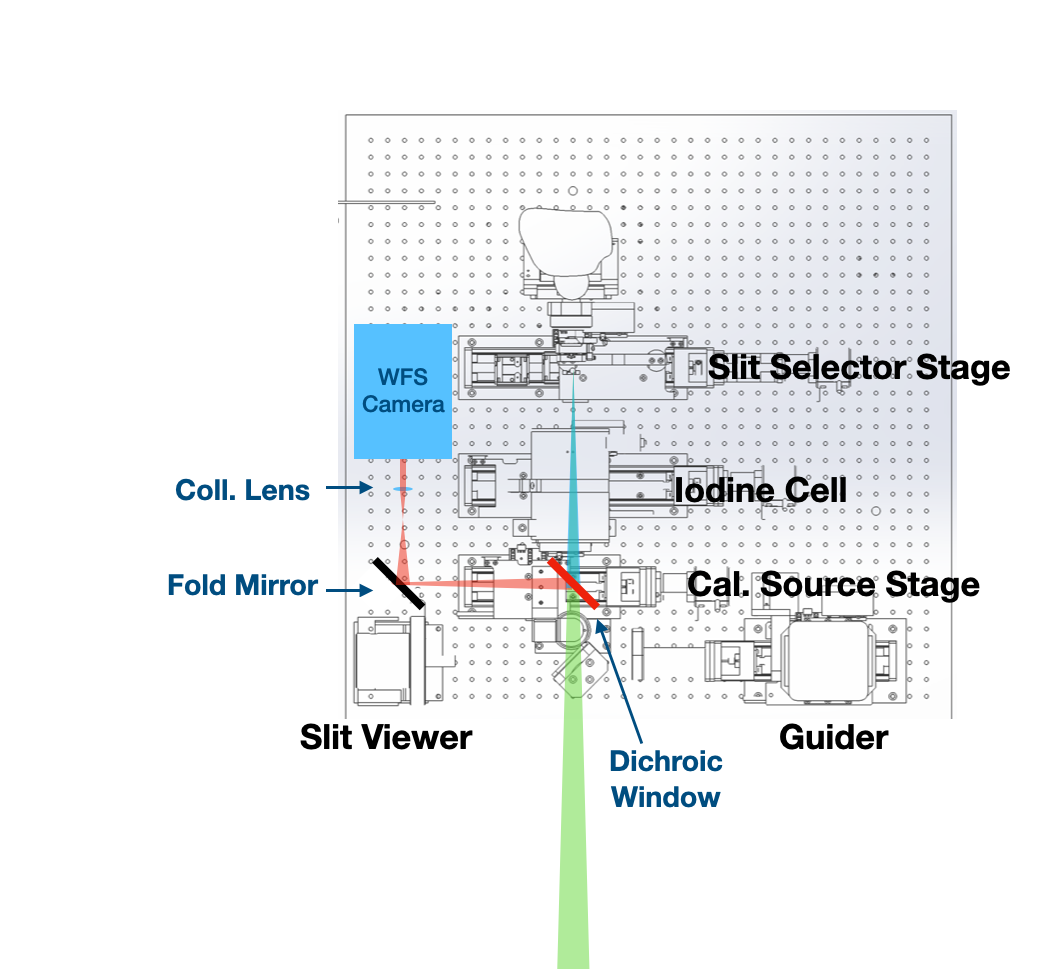}
   \end{tabular}
   \end{center}
   \caption[example] 
   { \label{fig:layout} 
 \textbf{The current APF optics layout with the proposed WFS Camera Location.} The current APF optics bench components are shown in black and white and the additions are shown in color. We will use the calibration source stage to allow insertion of a short-pass dichroic in the beam.  The dichroic will reflect long wavelength light to an Alpao SH WFS. A collimating lens is all that is needed to create a pupil image on the SH WFS camera. 
}
   \end{figure}

The WFS will be fed by a short pass dichroic that has a cut-on wavelength between 700-750 nm. The dichroic will feed a transmissive reimaging optic that images the APF pupil onto the Alpao Shack-Hartmann lenslet array. The fore optics for the Levy spectrometer are shown in Figure \ref{fig:layout}.  They are laid out on an optical breadboard that has suitable space for a dichroic, reimaging lens and the Alpao WFS module.

\begin{figure}
\centering
\includegraphics[width=1.0\linewidth]{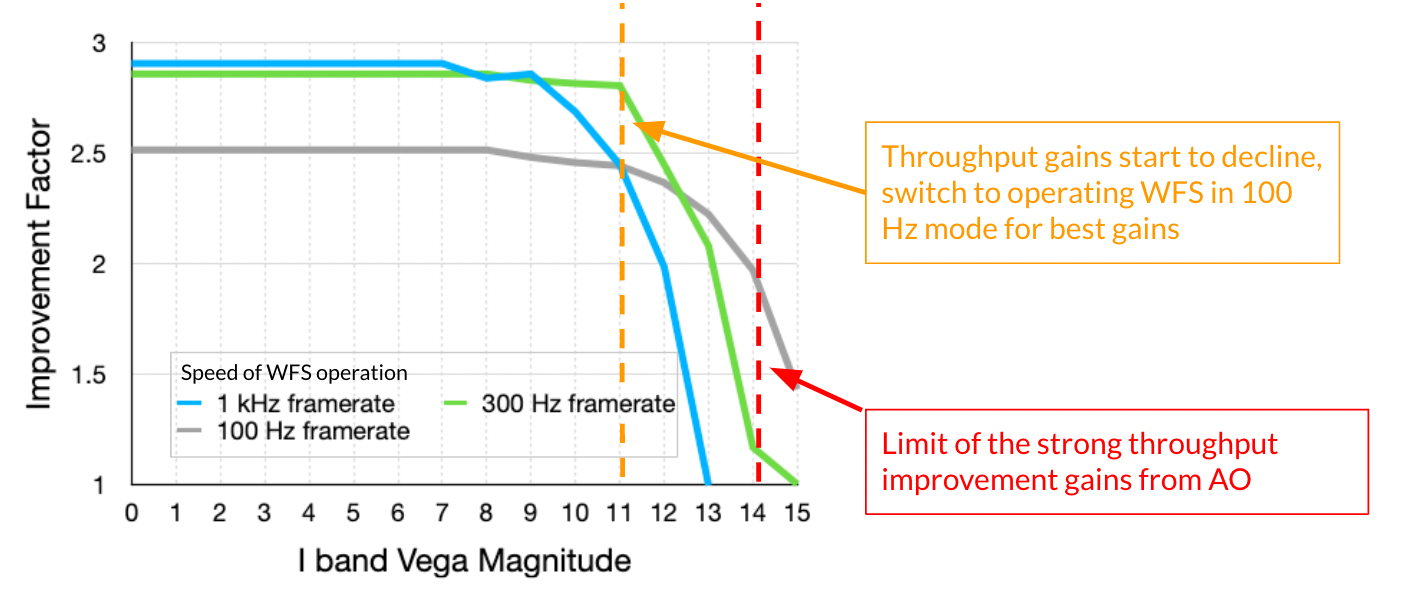}
 \caption[example]{\textbf{Simulated Efficiency Improvement of an APF enabled by AO.}  To understand what   throughput improvement we would expect from the AO system as we changed the brightness of the science target, we simulated the throughput improvement factor of the APF using HCIpy\cite{por2018hcipy}.  We predict that for targets brighter than an I-band magnitude of 11, the AO system can provide throughput gains to a factor of x$2.9$, when operating the WFS camera using a speed of 300Hz or 1kHz.  For targets with I-magnitude between 11 and 14, operating the WFS in 100 Hz mode will lead to improvement factors between 2 and 2.5.   \label{fig:IbandVega}}
   \end{figure}

The quantum efficiency for the Alpao EMCCD Shack-Hartmann WFS is  50\% (\href{https://www.alpao.com/wp-content/uploads/2021/04/WFS-datasheet-rev-2021a.pdf}{Alpao datasheet}).   Assuming a 30\% system throughput (not including the detector) with the specified noise for each subaperture from Alpao, we estimate being able to provide usable stabilization down to guide star magnitudes corresponding to an I-band magnitude of approximately I=11.  Simulations with the HCIpy software package\cite{Por2018} confirm this, indicating 1 kHz operation will be optimal to stars as faint as I=9, 300 Hz should be used for stars with 10-12, and that some improvement out to I=15 is possible with corrections at 100 Hz, as shown in Figure \ref{fig:IbandVega}. 

The WFS will be tested in the UCSC Lab for Adaptive Optics  prior to deployment on the APF. This testbed has a light source, turbulence generators, an Alpao DM97-15 deformable mirror, and locations for wavefront sensors to be tested.

\subsection{Integration into the APF}

Currently, the APF operates with a guider camera providing pointing information to the telescope at a rate of 1 Hz.  This system will be retained and used for an acquisition camera for the system.  Once the star is acquired on the WFS, the guiding corrections will be turned off and the AO loop will provide stabilization and 
pointing offsets for the telescope tracking software.  Since the acquisition will be carried out in the same way, we expect none to minor changes to the observing efficiency after the AO system is implemented. We expect to have a failsafe mode for seeing-limited operations, should conditions be too poor or variable for routine AO-corrected imaging on a particular night.

After validation testing, we will replace the existing static secondary mirror with the ASM on the APF.  The ASM will use the same focus and tip/tilt mechanism used for static positioning in the current APF. The initial tests at the APF will include 
measurement of the PSF in open loop, refinement of the secondary shape to optimize PSF, measurement of the closed loop performance,
 refinement of the automated performance, and integration operation into the queue scheduling.
 
 The real-time computer will be based on the Keck RTC setup developed for the infrared pyramid WFS on Keck.

\section{SIMULATED PERFORMANCE}

\subsection{What is currently limiting the performance of the APF?}

Analysis of guider data for a typical sequence allows us to estimate the slit throughput as 21\% with variations of 4\% as seen by the guider over 1 second integrations (see Figure \ref{fig:sim}). 
This throughput is limited by the spread of light because of atmospheric seeing and slow ($<$1 Hz) tracking variations. 

The time-varying line spread function is the largest source of systematic error in the radial velocity precision. 
When using the iodine cell method to measure radial velocities, the most critical component in the measurement accuracy is the line shape function~\cite{Butler1996}. The steeper the absorption features in the star, the more accurate the velocities. 
For the Levy spectrometer on the APF, the line shape function is modeled with 18 wavelength-dependent free parameters for every exposure. The classical solution to stabilizing the line spread function for an iodine cell spectrometer is retrofitting a fiber scrambling system~\cite{Spronck2013}. When implemented on the HIRES on the Keck I telescope, Spronck et al. (2015)\cite{Spronck2015} found that their system stabilized the line spread function to such a degree that the highest signal-to-noise data showed errors of 0.5 m/s, less than half of the best values of 1.1-1.2 m/s for the same signal to noise without the scrambling. Further, the line spread function was stable enough that it removed the need for the huge number of wavelength-dependent free parameters per observation. For the APF, Burt et al. (2014)\cite{Burt2014} found that the noise floor of the Levy spectrometer was 1.2 m/s, very close to the values found by Spronck et al. (2015)\cite{Spronck2015} for HIRES, showing that in principle a similar performance gain is possible. However, the key disadvantage of fiber scrambling is that the additional optics required meant that throughput had decreased by a factor of six. A similar result for the Levy on the APF would mean that the typical exposures would increase from 20 minutes to two hours. 

\subsection{Expected Image Stability and Slit Efficiency Improvement}

   \begin{figure}
   \begin{center}
   \begin{tabular}{l}
   \includegraphics[height=9.8cm]{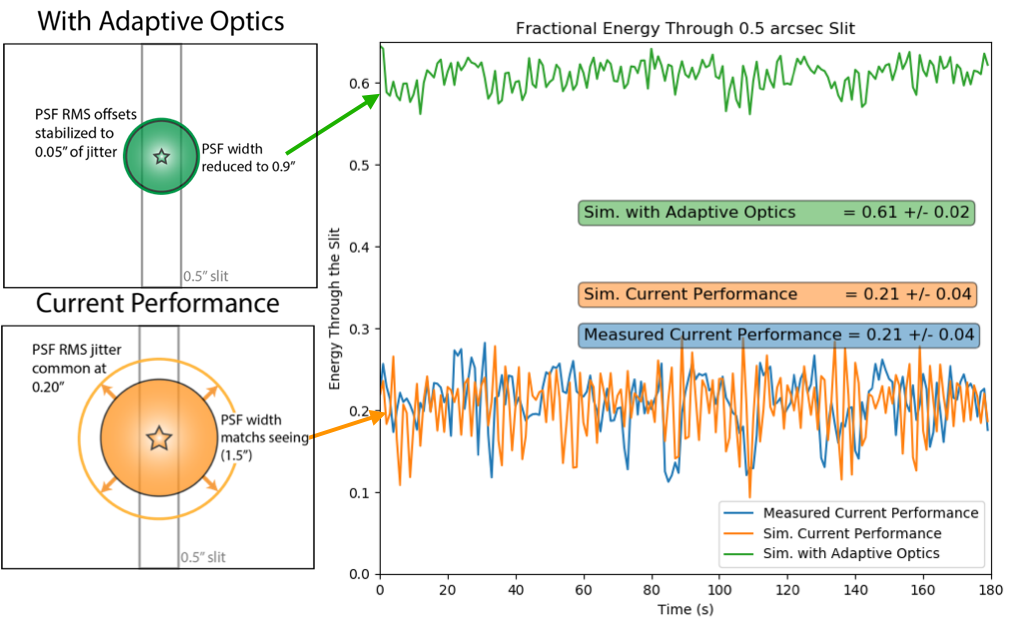}
   \end{tabular}
   \end{center}
   \caption[example] 
   { \label{fig:sim} 
\textbf{Simulated Improvements to Image Stablity and Throughput}. 
The APF currently allows for $\sim21 \pm 4$\% of the light through a 0.5" slit (blue). We simulated the current throughput and variability using
HCIpy~\cite{Por2018} (orange) and the modeled the system with AO correction (green). We find that the AO system reduces the PSF width by 40\%, leading to an expected improved throughput of $61 \pm 2$\% (green line). The RMS offset from the slit is improved by a factor of four resulting in better PSF stability.}
   \end{figure} 

AO correction is conventionally discussed in terms of diffraction-limited imaging, where correction on the spatial scale of the Fried length (r$_0$) is required to achieve a decent Strehl ratio. For the APF setup we have more modest goals: (1) improve the fraction of light making it through the slit, and (2) stabilize its variation.  
Qualitatively, this is similar to ground-layer AO (GLAO) correction, where correction of only the lowest order modes of the atmosphere that dominate the size of the image blur is needed to achieve improvements in the FWHM.  We note that other visible light systems have achieved this image sharpening in the visible (see, for example, Chun et al. (2018)~\cite{Chun2018}).  Thus the expected improvement is consistent with other AO systems, even if their technical approach (wide-field GLAO correction) is different.

To estimate the system performance, we simulated the expected improvement using the python package HCIpy~\cite{por2018hcipy} (see Figure \ref{fig:sim}). We created an atmosphere with similar image blur to guiding data taken from a typical observing sequence. We then added in slow tracking variations at several frequencies to match the variations we visually see in the guider.   The resulting slit efficiency (21\%) and variations of 4\%  are similar to what is seen by the APF guider. 

We then simulate closed-loop AO corrections for an AO system as described in Section 4.  
We find that the PSF width decreases from $1.5 \pm 0.2$ arcsec to $0.9 \pm 0.05$ arcsec in simulation in 1s snapshots of the simulation. The image jitter is reduced by a factor of four (from 0.20" to 0.05").
With the 0.5-arcsec slit width, this improved light concentration through the slit would lead to a throughput increase of $61\pm2\%$ for an improvement factor of 2.9. 
For the 1-arcsec slit width, the improvement is similar. Without AO correction the efficiency is 40\%, while after correction it is 79\% for a 2.0 times improvement factor.   

Even if only the efficiency improvement can be realized, the value of AO for the APF will increase the cadence of nightly observations.  The efficiency improvement suggests that we could be able to observe over double the number of stars per night compared to current observations.  We expect to provide this improvement on stars as faint as I=14 as shown in Figure \ref{fig:IbandVega}. This dramatic speed-up of observations is equivalent to having a second APF available.

\subsection{RV Measurement Precision}

The connection between PSF image stability and RV stability is complex and difficult to quantify. We do not attempt to  quantify the RV measurement gains, however, we note that there is evidence to suggest that improved image stabilization will indeed lead to higher precision RV measurements. 

One example of their correlation is seen in HARPS data where LoCurto et al. (2010)~\cite{LoCurto2010} notes that a guider variation of 0.1 arcsec results in approximately 0.3 m/s errors. While this example cannot be used as a direct comparison because HARPS is a different style spectrograph to APF/Levy, it demonstrates that RV and image stability are correlated. We expect the improved stability seen in our simulations to benefit the achieved precision improvement in a similar way to the fiber scrambler retrofit to HIRES\cite{Spronck2013} without the reduction in throughput. 

In total, we may expect to be able to get to the same signal-to-noise thresholds in less than half the time while also doubling the precision of the measurement due to the improvement in image stability.  Quantifying the improvement to RV accuracy using the APF AO system on-sky will be informative for future AO-enhanced RV spectrometer designs.

\acknowledgments 
The authors would like to thank Mark Chun, Daren Dillon, Wouter Jonker, Stefan Kuiper, Matthew Maniscalco, Claire Max, and the members of the Lab for Adaptive Optics at UC Santa Cruz for their guidance and assistance in the completion of this work.  

\bibliography{report} 
\bibliographystyle{spiebib} 

\end{document}